\documentclass[twocolumn,aps,showpacs,prl] {revtex4-1}
\usepackage{graphicx}
\usepackage{amssymb}
\usepackage{amsmath}
\usepackage{color}
\usepackage[utf8x]{inputenc}
\newcommand{\VEC}[1]{\mathbf{#1}}
\newcommand{\rvec}{\VEC{r}}
\newcommand{\kvec}{\VEC{k}}
\newcommand{\fig}[1]{Fig.~\ref{f:#1}}

\newcommand{\stext}[1]{\cite{SOM}}

\begin{document} 
\title{Two-step melting in two dimensions: First-order liquid-hexatic
transition} 
\author{Etienne P. Bernard}
\email{etienne.bernard@lps.ens.fr}
\author{Werner Krauth}
\email{werner.krauth@ens.fr}
\affiliation{
Laboratoire de Physique Statistique\\ Ecole Normale Supérieure,
UPMC, CNRS\\
24 rue Lhomond, 75231 Paris Cedex 05, France}

\date{}
\begin{abstract}
Melting in two spatial dimensions, as realized in thin films or at
interfaces, represents one of the most fascinating phase transitions
in nature, but it remains poorly understood.  Even for the fundamental
hard-disk model, the melting mechanism has not been agreed on
after fifty years of studies. A recent Monte Carlo algorithm allows us to
thermalize systems large enough to access the 
thermodynamic regime. We show that melting in hard disks 
proceeds in two steps with a liquid phase, a hexatic phase, and a 
solid. The hexatic-solid transition is continuous while, 
surprisingly, the liquid-hexatic transition is of first-order.
This melting scenario solves one of the fundamental statistical-physics 
models, which is at the root of a large body of theoretical, computational 
and experimental research.
\end{abstract}
\maketitle

Generic two-dimensional particle systems cannot crystallize at finite 
temperature\cite{Peierls_34,Peierls_35,Mermin_Wagner} 
because of the importance of fluctuations, yet they may form 
solids\cite{Alder_62}.  
This paradox has provided the motivation for
elucidating the fundamental melting transition in 
two spatial dimensions. A crystal is characterized by particle positions 
which fluctuate about the sites of an infinite regular lattice. It has
long-range \emph{positional} order. Bond orientations 
are also the same throughout the lattice. A
crystal thus possesses long-range \emph{orientational} order. 
The positional correlations of a  
two-dimensional solid decay to zero as a power law at large distances.
Because of the absence of a scale, one speaks of  
``quasi-long range'' order. In a two-dimensional solid, the 
lattice distortions preserve long-range orientational order\cite{Mermin}, while in  
a liquid, both the positional and the orientational correlations decay exponentially.

Besides the solid and the liquid, a third phase, called ``hexatic'',
has been discussed but never clearly identified in particle systems.
The hexatic phase is characterized by exponential positional but
quasi-long range orientational correlations. It has long been  
discussed whether the melting transition follows a one-step
first-order scenario between the liquid and the solid (without the
hexatic) as in three spatial dimensions\cite{Hoover_Ree}), or whether it agrees
with the celebrated Kosterlitz, Thouless\cite{Kosterlitz_Thouless},
Halperin, Nelson\cite{Halperin_Nelson} and Young\cite{Young} (KTHNY)
two-step scenario with a hexatic phase separated by continuous
transitions from the liquid and the solid\cite{Strandburg, Lee_Strandburg,
Zollweg_Chester, Weber_Marx, Jaster, Mak, Zahn, Bagchi, Peng}.

Two-dimensional melting was discovered \cite{Alder_62} in the simplest 
particle system, the hard-disk model. Hard disks (of radius 
$\sigma$) are structureless and all configurations of non-overlapping 
disks have zero potential energy. Two isolated disks only feel the 
hard-core repulsion, but the other disks mediate an entropic 
``depletion'' interaction (see, e.g., \cite{SMAC}). 
Phase transitions result from an ``order from disorder'' phenomenon: At high density, 
ordered configurations can allow for larger local fluctuations, thus
higher entropy, than the disordered liquid. For hard 
disks, no difference exists between the liquid and the gas. At fixed 
density $\eta$, the phase diagram is independent of temperature 
$T=1/k_\text{B}\beta$, and the pressure is proportional to $T$, as 
discovered by D. Bernoulli in 1738. Even for this basic model, the nature 
of the melting transition has not been agreed on. 

The hard-disk model has been simulated with the local Monte Carlo 
algorithm since the original work by Metropolis et al. \cite{Metropolis}.
A faster collective-move ``event-chain'' Monte Carlo algorithm 
was developed only recently\cite{Bernard} 
(see \cite{SOM}).
We will use it to show that the melting transition  neither 
follows the one-step first-order nor the two-step continuous KTHNY scenario.  

\begin{figure*}[ht!]
\includegraphics[width=0.8\textwidth]{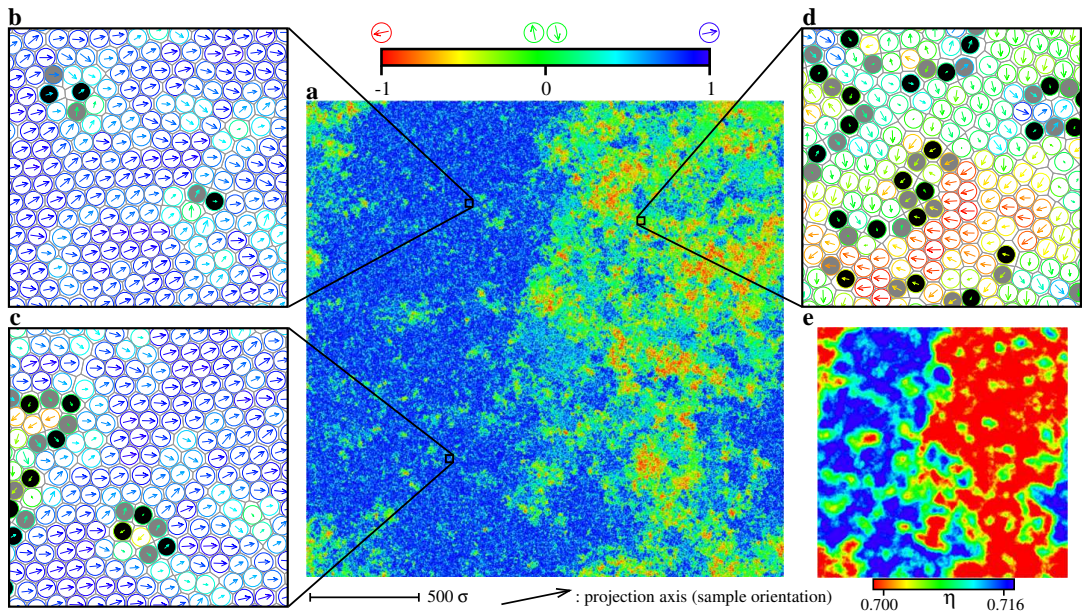}
\caption{
   Phase-coexistence for $1024^2$ thermalized hard disks at
   density $\eta=0.708$. \textbf{a}: Color-coded
   local orientations $\Psi_k$ showing long
   orientational correlations (blue region, see \textbf{b},\textbf{c})
   coexisting with short-range correlations (see \textbf{d}). 
   \textbf{e}: Local densities (averaged over a radius of $50\sigma$), demonstrating
   the connection between density and local orientation
   (see \cite{SOM}). In \textbf{b}, \textbf{c}, \textbf{d}, disks with 
   five (seven) neighbors are colored in gray (black).}
   \label{f:fig1}
\end{figure*}

To quantify orientational order, we express the local orientation of disk $k$
through the complex vector $\Psi_k = \langle \exp(6i\phi_{kl})
\rangle$, with $\langle \rangle$ the average over all the
neighbors $l$ of $k$. The angle $\phi_{kl}$ describes the 
orientation of the bond $kl$ with respect to a fixed axis. The sample 
orientation is defined as $\Psi=1/N \sum_k \Psi_k$. For a perfect triangular
lattice, all the angles $6 \phi_{kl}$ are the same and $|\Psi_k|=|\Psi|=1$
(see \cite{SOM}).

\begin{figure}[ht!]
   \includegraphics[width=0.5\textwidth]{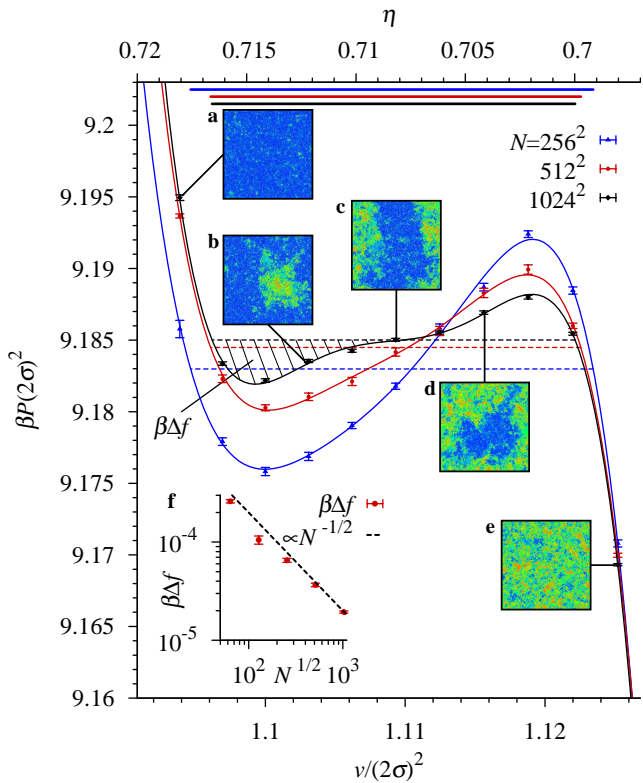}
   \caption{
   Equilibrium equation of state for hard disks.
   The pressure is plotted \emph{vs.} volume per particle
   ($v=V/N$) (lower scale) and density $\eta$ (upper scale)).  In the
   coexistence region, the strong system-size dependence stems from the
   interface free energy. The Maxwell constructions (horizontal lines) 
   suppress the interface effects (with a convex free energy) 
   for each $N$. ``Stripe''
   (\textbf{c}, for $N=1024^2$) and ``bubble'' configurations
   (\textbf{b},\textbf{d}) are shown in the coexistence region,
   together with two single-phase configurations (\textbf{a},\textbf{e}).
   The interface free energy per disk $\beta\Delta f$ 
   (hatched area) scales as $1/\sqrt{N}$ (\textbf{f}).} 
   \label{f:equation_of_state}
\end{figure}

In \fig{fig1}, the local orientations of a configuration with $N=1024^2$ 
disks at density $\eta=N\pi\sigma^2/V=0.708$ in a square box of volume $V$ 
are projected onto the sample orientation and represented using a color 
code (see \cite{SOM}).
Inside this configuration, a vertical 
stripe with density $\sim 0.716$ preserving orientational 
order over long distances coexists with a stripe of disordered liquid of 
lower density $\sim 0.700$. Each stripe corresponds to a different phase.
The two interfaces of length $\simeq \sqrt{N}$ close on themselves 
\emph{via} the periodic boundary conditions. Stripe-shaped phases as in 
\fig{fig1}\textbf{a} 
are found in the center of a coexistence 
interval $\eta \in [0.700,0.716]$, whereas close to its endpoints, a 
``bubble'' of the minority phase is present inside the majority phase 
for $\eta\gtrsim 0.700$ and $\eta \lesssim 0.716$ (see \fig{equation_of_state}).
This phase coexistence is the hallmark of a first-order transition.

The first-order transition shows up in the equilibrium equation of state 
$P(V)$ (see \fig{equation_of_state}). At finite $N$, the free energy is not necessarily 
convex (as it would be in an infinite system) and the equilibrium pressure 
$P(V)  = - \partial F/\partial V$ can form a thermodynamically stable 
loop due to the interface free energy. 
The pressure loop in the coexistence window of a 
finite system is caused by
the curved  interface between a bubble of minority phase and the surrounding majority 
phase (see \fig{equation_of_state}\textbf{b},\textbf{d})).
In a system with periodic boundary conditions, the pressure loop contains 
a  horizontal piece corresponding  to the ``stripe'' regime, where the 
interfaces are flat. This is visible near $\eta \sim 0.708$ for the largest systems 
in \fig{equation_of_state}. 
In a finite system, the Maxwell construction suppresses the interface effects.
For the equation of state of \fig{equation_of_state}\textbf{a}, this construction
confirms the boundary densities $\eta=0.700$ and $\eta=0.716$ of \fig{fig1} 
for the coexistence interval, with very small finite-size effects.
The interface free energy per disk, the hatched area in 
\fig{equation_of_state}, depends on the length $\propto \sqrt{N}$ of the 
interface in the ``stripe'' regime 
so that $\Delta f = \Delta F/N \propto 1/\sqrt{N}$ (see \fig{equation_of_state}\textbf{f}).

The first-order nature of the transition involving the liquid is thus established by 
\emph{i)}: The visual evidence of phase coexistence in \fig{fig1}, 
\emph{ii)}:  The $\propto 1/\sqrt{N}$  scaling of the interface free energy 
per disk\cite{Lee_Kosterlitz}, and 
\emph{iii)}: The characteristic shape of the 
equation of state in a finite periodic system 
\cite{Wood_2_d, Binder_Method,Schrader_Binder}. 
We stress that the system size is larger than the physical 
length scales so that the results hold in the thermodynamic limit (see \cite{SOM}).

\begin{figure*}[ht!]
  \includegraphics[width=0.8\textwidth]
   {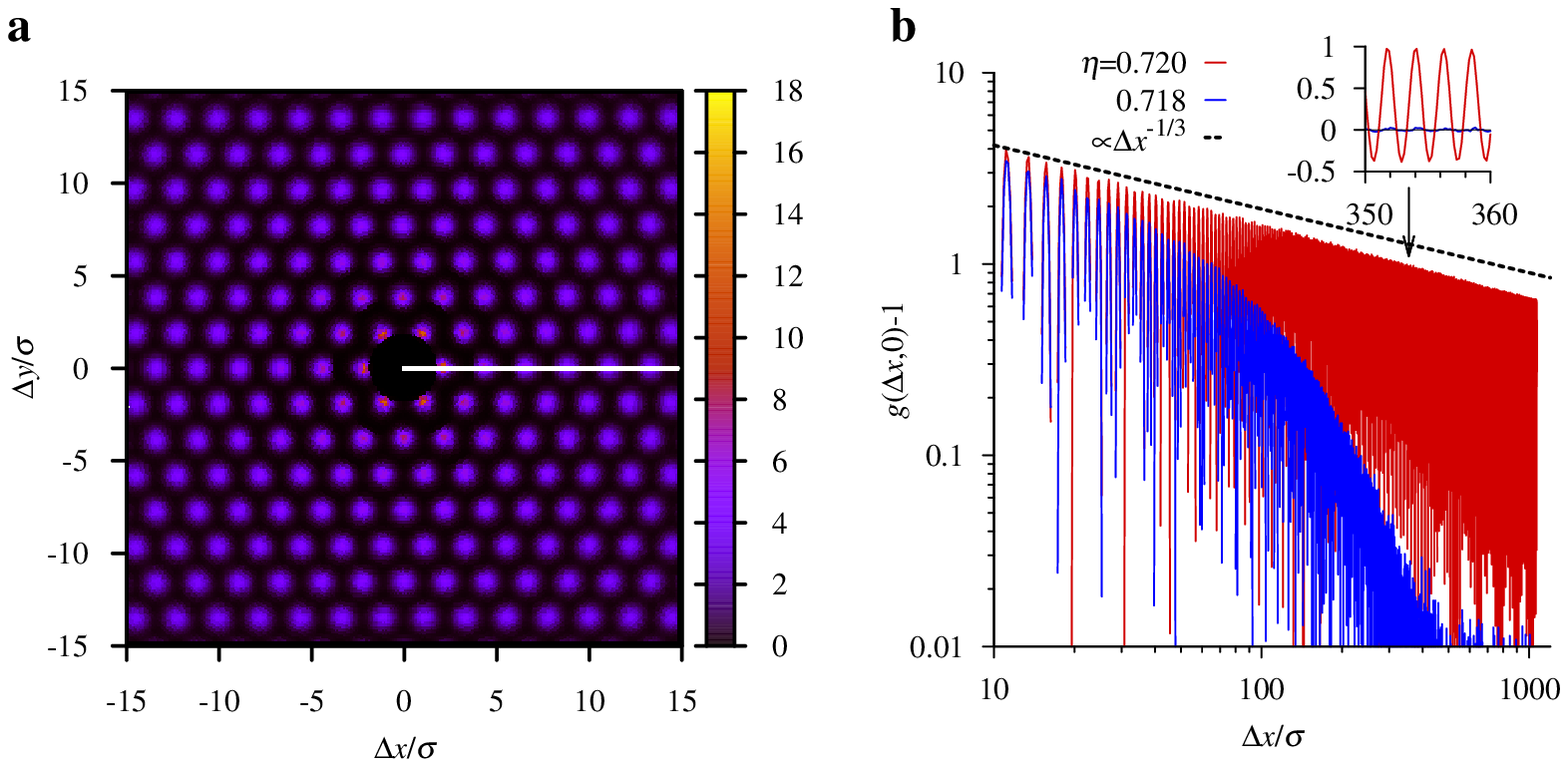}
  \caption{
   Configuration-averaged two-dimensional pair correlation.
   $g(\Delta
   \rvec)$ is obtained from the two-dimensional histogram of periodic
   distances $\Delta \rvec_{ij} = \rvec_i-\rvec_j$.  \textbf{a}: Pair
   correlation $g(\Delta \rvec)$ at density $\eta=0.718$ for small
   $\Delta \rvec=(\Delta x,\Delta y)$.  Each disk configuration is
   oriented with respect to $\Psi$.  The excellent contrast between the
   peak and the bottom values of $g(\Delta \rvec)$ at $|\Delta \rvec|
   \gtrsim 2 \sigma$, of about $(16: 0.2)$, provides evidence for
   the single-phase nature of the system.  \textbf{b}: Cut of the 
   sample-averaged
   $g(\Delta \rvec)-1$ for $\Delta \rvec=(\Delta x,0)$. Decay
   is exponential for $\eta=0.718$ and algebraic
   for $\eta = 0.720$. (See \cite{SOM}
   for positional and orientational correlation functions.)}
   \label{f:density_correlations_averaged}
\end{figure*}

In the coexistence interval, the individual phases are difficult to analyze
at large length scales because of the fluctuating interface, 
and only the low-density coexisting phase is identified as a liquid
with orientational correlations below a scale of $\sim 100\sigma$ (see 
\fig{fig1}\textbf{a},\textbf{d}). Unlike constant $NV$ 
simulations, Gibbs ensemble simulations can have phase coexistence without 
interfaces, but these simulations are very slow at large $N$ (see \cite{SOM}).
The single-phase system at density $\eta=0.718$, is above the coexistence 
window for all $N$ (see \fig{equation_of_state}), and it allows us to 
characterize the high-density coexisting phase.

Positional order can be studied in the two-dimensional pair
correlation $g(\Delta \rvec)$, the high-resolution histogram of
periodic pair distances $\Delta \rvec_{ij} = \rvec_i-\rvec_j$ sampled from all 
$ N(N-1)/2$ pairs $i,j$ of disks.
To average this two-dimensional histogram over configurations 
(as in \fig{density_correlations_averaged}) the latter are 
oriented such that the $\Delta x$ axis points in the direction of the sample 
orientation $\Psi$. 
At short distances, hexagonal order is evident at $\eta=0.718$
(see \fig{density_correlations_averaged}\textbf{a}). The excellent contrast 
between peaks and valleys of $g(\Delta \rvec)$ at small 
$|\Delta \rvec| \gtrsim 2 \sigma$ underlines
the single-phase nature of the system at this density.
The cut of the histogram along the positive $\Delta x$ axis leaves no 
doubt that the system has exponentially decaying positional order on a 
length scale of $\sim 100 \sigma$ and cannot 
be a solid. The (one-dimensional) positional correlation function 
$c_\kvec(r)$, computed by Fourier transform of $g(\Delta \rvec)$,  
fully confirms these statements (see \cite{SOM}).

The orientational correlations at density $\eta=0.718$ decay extremely 
slowly and do not allow us to distinguish between quasi-long range 
and long-range order (see \cite{SOM}). However, short-ranged positional 
correlation is inconsistent with
long-ranged orientational order. It follows that the 
orientation must be quasi-long ranged with a small exponent $\lesssim 0$, 
and that the system at $\eta=0.718$ and  the high-density coexisting phase
are both hexatic.

The two-dimensional pair correlation $g(\Delta \rvec)-1$ of
\fig{density_correlations_averaged}\textbf{b} allows us to follow the transition 
from the hexatic to the solid: The positional order
increases continuously with density and 
crosses over into power-law behavior at density $\eta
\sim 0.720$, with an exponent $\simeq -1/3$ which corresponds  
to the stability limit of the solid phase in the 
KTHNY scenario. The hexatic-solid transition thus takes place 
at $\eta \gtrsim 0.720$. At this density,
the positional correlation function at large distances $r$,
displays the finite-size effects characteristic
of a continuous transition, but up to a few hundred $\sigma$, $c_\kvec$ is
well stabilized with system size (see \cite{SOM}).
Moreover, no pressure loop is observed in the equation of state, and the
compressibility remains very small. The system is clearly in 
a single phase. Unlike the liquid-hexatic transition,
the hexatic-solid transition therefore follows the KTHNY scenario, and is
continuous.

The single-phase hexatic regime is confined to a density interval 
$\eta \in [0.716,0.720]$. Although narrow, it is an order of magnitude 
larger than the scale set by density fluctuations 
for our largest systems and can be be easily resolved
(see \cite{SOM}). In the hexatic phase, the
orientational correlations decay extremely slowly. The exponent of 
the orientational correlations  is close to zero and negative. 
It remains far from the lower limit of
$-1/4$ at the continuous KTHNY transition, as this transition is
preempted by a first-order instability.

\begin{figure*}[ht!]
   \includegraphics[width=0.8\textwidth]{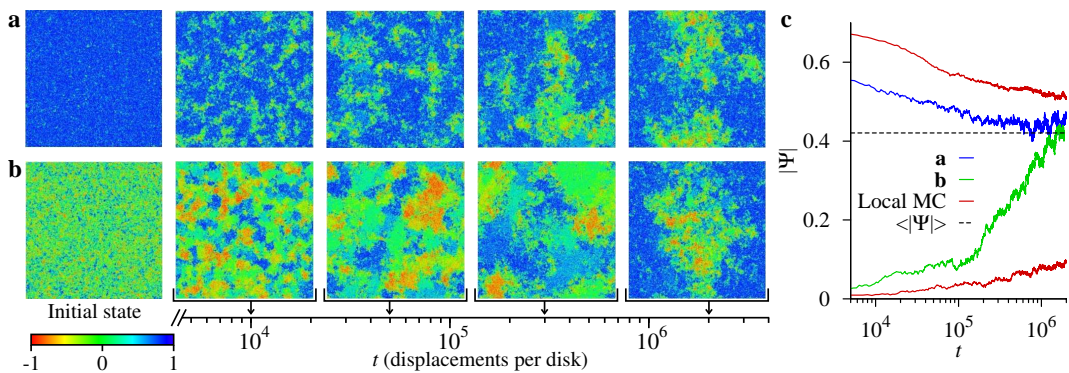}
   \caption{
   Approach to thermal equilibrium from different initial
   conditions. \textbf{a},\textbf{b}: $1024^2$ hard disks at density
   $\eta=0.708$, after a quench from a high-density crystal (\textbf{a})
   and from a low-density liquid (\textbf{b}), showing coarsening
   leading to phase separation (Color code for $\Psi_k$  as in \fig{fig1}\textbf{b}, 
   see also \cite{SOM}).
   Each of the runs takes about one week
   of CPU time. \textbf{c}: Absolute value of the sample orientation for
   the simulations in \textbf{a},\textbf{b}, compared to runs with
   the local Monte Carlo algorithm from the same initial conditions
   (time in attempted displacements per disk). The correlation time
   of the event-chain algorithm, on the order of $10^6$ displacements
   per disk, estimated from \textbf{c}, agrees with the correlation time
   estimated in our production runs with $6\times 10^7$ total 
   displacements per disk.}
   \label{f:Krauth_fig_4}
\end{figure*}

The event-chain algorithm is about two orders of magnitude faster than
the local Monte Carlo used up to now, allowing us to thermalize for the
first time dense systems with up to $1024^2$ disks. To illustrate convergence
toward thermal equilibrium and to check that hard disks in the window
of densities $\eta \in [0.700,0.716]$ are indeed phase-separated,
we show in \fig{Krauth_fig_4} two one-week simulations of our
largest systems after quenches from radically different initial
conditions, namely the (unstable) crystal, with $|\Psi|=1$, and the
liquid, for which $|\Psi|\simeq 0$.  For both initial conditions, 
a slow process of coarsening takes place
(see \fig{Krauth_fig_4}\textbf{a},\textbf{b}). 
Phase separation is observed after $ \sim 10^6$ displacements per
disk, and the sample orientation takes on similar absolute values
(see \fig{Krauth_fig_4}\textbf{c}). Effective simulation times of many earlier
calculations were much shorter\cite{Jaster,Mak}, and the simulations remained in an
out-of-equilibrium state which is homogeneous on large length 
scales, whereas the thermalized system is phase-separated and therefore 
inhomogeneous. The
production runs for $N=1024^2$ were obtained from Markov chains with 
running times of nine months, $30$ times larger than those 
of \fig{Krauth_fig_4}\textbf{a},\textbf{b}.

The solution of the melting problem presented in this work
provides the starting point for the understanding of melting in 
films, suspensions, and other soft-condensed-matter systems. The 
insights obtained combine thermodynamic reasoning  with powerful tools: 
advanced simulation algorithms, direct visualization, and a failsafe 
analysis of correlations. These tools will all be widely applicable, for 
example to study the cross-over from two to three-dimensional melting    
as it is realized experimentally with spheres under different 
confinement conditions\cite{Peng}.

In simple systems such as hard disks and spheres, entropic and elastic effects
have the same origin: elastic forces are entropically induced. 
For general interaction potentials, entropy and 
elasticity are no longer strictly linked and order-disorder 
transitions, which can then take place as a function of temperature 
or of density, might realize other melting scenarios\cite{Enter}.
Theoretical, computational  and experimental research on more complex 
microscopic models will build on the hard-disk solution obtained in this 
work.
\begin{acknowledgments}
We are indebted to  K. Binder and D. R. Nelson  for helpful 
discussions and correspondence. We thank J. Dalibard and G. Bastard for 
a critical reading of the manuscript.
\end{acknowledgments}

\end{document}